\documentclass[usenatbib]{mn2e}
\usepackage{graphicx,times}
\usepackage{bm,url}
\graphicspath{{./fig/}{./png/}}
\volume{422}
\pagerange{2465--2473}
\pubyear{2012}

\newcommand{\EQ}{\begin{equation}}
\newcommand{\EN}{\end{equation}}
\newcommand{\EQA}{\begin{eqnarray}}
\newcommand{\ENA}{\end{eqnarray}}

\newcommand{\Eq}[1]{Eq.~(\ref{#1})}

\newcommand{\Sec}[1]{Sect.~\ref{#1}}

\newcommand{\Fig}[1]{Fig.~\ref{#1}}
\newcommand{\FFig}[1]{Figure~\ref{#1}}
\newcommand{\Figs}[2]{Figs.~\ref{#1} and \ref{#2}}
\newcommand{\Figss}[2]{Figs.~\ref{#1}--\ref{#2}}

\newcommand{\mean}[1]{\overline #1}
\newcommand{\meanv}[1]{\bm{\overline #1}}
\newcommand{\pd}{\partial}

\newcommand{\meanrho}{\overline{\rho}}

{}
{}
\newcommand{\meanFFFF}{\overline{\mbox{\boldmath ${\cal F}$}}{}}{}

\newcommand{\meanSSSS}{\overline{\mbox{\boldmath ${\mathsf S}$}} {}}

{}
{}
{}
{}
{}
{}
{}
\newcommand{\meanAA}{\overline{\mbox{\boldmath $A$}}{}}{}
\newcommand{\meanBB}{\overline{\mbox{\boldmath $B$}}{}}{}
{}
{}
{}
{}
{}
\newcommand{\meanJJ}{\overline{\mbox{\boldmath $J$}}{}}{}
\newcommand{\meanUU}{\overline{\bm{U}}}

{}
{}
{}

\newcommand{\meanB}{\overline{B}}

\newcommand{\meanU}{\overline{U}}

\newcommand{\means}{\overline{s}}
\newcommand{\meanp}{\overline{p}}

\newcommand{\meanT}{\overline{T}}

\newcommand{\meanFFF}{\overline{\cal F}}

\newcommand{\hatBB}{\hat{\bm{B}}}
\newcommand{\BBeta}{\hat{\bm{B}}} 
{}

{}
{}

\newcommand{\Beta}{\hat{B}}  

\newcommand{\FF}{\bm{F}}

\newcommand{\uu}{\mbox{\boldmath $u$} {}}
\newcommand{\UU}{\mbox{\boldmath $U$} {}}

\newcommand{\bb}{\mbox{\boldmath $b$} {}}
\newcommand{\BB}{\mbox{\boldmath $B$} {}}

\newcommand{\grav}{\mbox{\boldmath $g$} {}}
\newcommand{\nab}{\mbox{\boldmath $\nabla$} {}}

\newcommand{\DD}{{\rm D} {}}

\newcommand{\dd}{{\rm d} {}}

\def\la{\mathrel{\mathchoice {\vcenter{\offinterlineskip\halign{\hfil
$\displaystyle##$\hfil\cr<\cr\sim\cr}}}
{\vcenter{\offinterlineskip\halign{\hfil$\textstyle##$\hfil\cr<\cr\sim\cr}}}
{\vcenter{\offinterlineskip\halign{\hfil$\scriptstyle##$\hfil\cr<\cr\sim\cr}}}
{\vcenter{\offinterlineskip\halign{\hfil$\scriptscriptstyle##$\hfil\cr<\cr\sim\cr}}}}}
\def\ga{\mathrel{\mathchoice {\vcenter{\offinterlineskip\halign{\hfil
$\displaystyle##$\hfil\cr>\cr\sim\cr}}}
{\vcenter{\offinterlineskip\halign{\hfil$\textstyle##$\hfil\cr>\cr\sim\cr}}}
{\vcenter{\offinterlineskip\halign{\hfil$\scriptstyle##$\hfil\cr>\cr\sim\cr}}}
{\vcenter{\offinterlineskip\halign{\hfil$\scriptscriptstyle##$\hfil\cr>\cr\sim\cr}}}}}
\def\Ra{\mbox{\rm Ra}}
\def\Ma{\mbox{\rm Ma}}

\def\Rem{\mbox{\rm Rm}}
\def\Pm{{\rm Pm}}
\def\Rm{R_{\rm m}}

\def\Rey{\mbox{\rm Re}}

\def\betap{\beta_{\rm p}}

\def\betag{\beta_{\rm g}}
\def\Bp{B_{\rm p}}
\def\Bs{B_{\rm s}}
\def\Bg{B_{\rm g}}
\def\qpz{q_{\rm p0}}
\def\qsz{q_{\rm s0}}
\def\qgz{q_{\rm g0}}
\def\qez{q_{\rm g0}}
\def\qp{q_{\rm p}}
\def\qs{q_{\rm s}}
\def\qg{q_{\rm g}}
\def\qe{q_{\rm g}}

\def\zinfty{z_\infty}
\def\csz{c_{\rm s0}}
\def\cp{c_{\rm p}}
\def\cs{c_{\rm s}}

\def\kf{k_{f}}

\def\Peff{{\cal P}_{\rm eff}}

\def\urms{u_{\rm rms}}

\def\nuT{\nu_{\rm T}}
\def\nut{\nu_{\rm t}}

\def\chit{\chi_{\rm t}}

\def\etaT{\eta_{\rm T}}
\def\etat{\eta_{\rm t}}

\def\Beqz{B_{\rm eq0}}

\def\Beq{B_{\rm eq}}

\def\half{{\textstyle{1\over2}}}

\def\onethird{{\textstyle{1\over3}}}

\newcommand{\yapj}[3]{ #1, {ApJ,} {#2}, #3}

\newcommand{\yapjl}[3]{ #1, {ApJ,} {#2}, #3}

\newcommand{\yan}[3]{ #1, {Astron.\ Nachr.,} {#2}, #3}

\newcommand{\yana}[3]{ #1, {A\&A,} {#2}, #3}

\newcommand{\ysovl}[3]{ #1, {Sov.\ Astron.\ Lett.,} {#2}, #3}
\newcommand{\yjetp}[3]{ #1, {Sov.\ Phys.\ JETP,} {#2}, #3}

\newcommand{\ymn}[3]{ #1, {MNRAS,} {#2}, #3}
\newcommand{\ynat}[3]{ #1, {Nature,} {#2}, #3}

\newcommand{\ysph}[3]{ #1, {Solar Phys.,} {#2}, #3}

\newcommand{\ypre}[3]{ #1, {Phys.\ Rev.\ E,} {#2}, #3}

\newcommand{\ybook}[3]{ #1, {#2} (#3)}

\newcommand{\psph}[1]{ #1, {Solar Phys.}, to be published}

\title[
Negative magnetic pressure in convection
]{
Negative effective magnetic pressure in turbulent convection
}

\author[P.\ J.\ K\"apyl\"a et al.]{
P.\ J.\ K\"apyl\"a$^{1,2}$,
A.\ Brandenburg$^{2,3}$,
N.\ Kleeorin$^{4,2}$,
M.\ J.\ Mantere$^{1}$, and
I.\ Rogachevskii$^{4,2}$
\\
$^1$Department of Physics, Gustaf H\"allstr\"omin katu
2a (PO Box 64),
FI-00064 University of Helsinki, Finland\\
$^2$NORDITA, AlbaNova University Center, Roslagstullsbacken 23,
SE-10691 Stockholm, Sweden\\
$^3$Department of Astronomy, AlbaNova University Center,
Stockholm University, SE-10691 Stockholm, Sweden\\
$^4$Department of Mechanical Engineering, Ben-Gurion University
of the Negev, PO Box 653, Beer-Sheva 84105, Israel\\
}

\date{Accepted 2012 February 22. Received 2012 February 21; in original form 2011 April 23}

\begin{document}
\maketitle

\begin{abstract}
  We investigate the effects of weakly and strongly stratified
  turbulent convection on the mean effective Lorentz force,
  and especially on the mean effective magnetic pressure.
  Earlier studies with isotropically forced
  non-stratified and stratified
  turbulence have shown that the contribution
  of the turbulence to the mean magnetic pressure is negative
  for mean horizontal magnetic fields that are smaller than the equipartition
  strength, so that the effective mean magnetic pressure
  that takes into account the turbulence effects, can be negative.
  Compared with earlier cases of forced turbulence
  with an isothermal equation of state,
  we find that the turbulence effect is similar to or
  even stronger in the present case of turbulent convection.
  This is argued to be due to the anisotropy
  of turbulence in the vertical direction.
  Another important difference compared with earlier studies
  is the presence of an evolution equation for the specific entropy.
  Mean-field modelling with entropy evolution indicates that the negative
  effective magnetic pressure can still lead to a large-scale
  instability which forms local flux concentrations, even though
  the specific entropy evolution tends to have a stabilizing effect
  when applied to a stably stratified (e.g., isothermal) layer.
  It is argued that this large-scale instability could be important for the
  formation of solar large-scale magnetic structures such as active regions.
\end{abstract}

\label{firstpage}

\section{Introduction}
The spatial and temporal coherence of the large-scale magnetic
field of
the Sun is manifested by sunspots appearing
within a certain range of latitudes from one cycle to the next.
A hydromagnetic dynamo is commonly held responsible
for the generation and maintenance of
large-scale magnetic fields \citep[cf.][]{M78,P79,KR80,RH04,BS05}.
Some models \citep{P55,P82,P84,SW80,S81,SCM94,DC99}, known as
flux transport dynamos, rely on the existence of strong
magnetic flux tubes at the base of the convection zone
or somewhat below, see also reviews
by \cite{Hugh07} and \cite{Tob07}.
These concentrations of magnetic field are thought to become
unstable once the field strength exceeds a critical value. The
subsequent rise of magnetic flux tubes to the surface
is used to explain active regions and sunspots.
Such models, however, face a number of
serious issues: firstly, the required strength of the magnetic
flux tubes is of the order of $10^5$ gauss \citep{DSC93},
which is expected to be a hundred times the equipartition
strength which is at odds with estimates that
the tachocline becomes
unstable already when fields of the order
of $10^3$ gauss are present \citep[e.g.][]{ASR05}.
Such strong fields are also hard to produce by
a turbulent dynamo \citep{GK11}.

An alternative scenario for the large-scale solar
magnetic field is
that it is maintained within the convection zone
by a distributed
dynamo, which generates diffuse sub-equipartition
strength magnetic fields \citep[e.g.][]{S76,B05,KKT06}.
Unlike the flux-transport dynamo, a distributed dynamo
does not directly explain the existence of sunspots
and active regions.
The alternative idea that sunspots have their origin
within the convection zone and that they might thus be
shallow phenomena is supported by observations
showing that the rotation rate of the Sun,
as measured by sunspots, depends monotonically on
their age so that young spots are the fastest and
oldest spots are the slowest \citep{PT98,B05}.
If one imagines sunspots floating in the plasma,
the rotation rate of the
youngest spots corresponds to roughly that at $r=0.95R_\odot$.
The decreasing rotation rate as a function of age
is consistent with older
spots being anchored at increasingly higher layers
where $\Omega$ is smaller due to its negative radial gradient near the
surface; see Fig.~4 of \cite{bene}.
This suggests that sunspots may
form near the surface of the Sun rather
than through the buoyant
rise of coherent flux tubes from the tachocline.
This is therefore compatible
with the distributed dynamo picture provided the
diffuse fields
within the convection zone can form concentrations like
sunspots near the surface \citep{B05}.

A promising mechanism that can form strong concentrations
from an initially uniform magnetic field was suggested
by \cite{KRR89,KRR90} who considered
the effects of turbulence or turbulent convection
on the large-scale Lorentz-force.
This work has been elaborated upon in a number of subsequent papers
\citep{KR94,KMR96,RK07}.
They find that, for a given range of large-scale magnetic field strengths,
there is a negative turbulence contribution to the mean magnetic pressure,
and the effective mean magnetic pressure
that accounts for the turbulence effects
can be negative.
This results in an excitation of
a large-scale instability.
The growth rate of the instability increases as
a function of density stratification \citep{KBKR11}.
In the Sun the density drops steeply
in the outermost layers, which favours the development of this
instability there.

The strongly stratified large-eddy simulations of
\cite{U09} and \cite{Kiti10}
may already have detected magnetic flux concentrations in turbulent
convection formed from initially uniform vertical magnetic fields.
Note also that a segregation into strongly and
weakly magnetized regions in magneto-convection
has been observed in numerical simulations at large
aspect ratios by \citep{Tao98,TP12}, which may have its origin
in some mean-field effect of the type considered here.
Also simulations of \cite{Stein_etal11} with a horizontal uniform field
at the bottom of the domain show emergence of magnetic flux structures,
while a number of numerical studies
\citep[e.g.][]{Schus06,Mart08,Remp09}
use strongly nonuniform fields as initial or boundary conditions.
The origin of such nonuniform fields is therefore not addressed
in these latter studies.

Direct numerical simulations (DNS) of homogeneous
\citep[][hereafter BKR]{BKR10} and density stratified \citep[][hereafter BKKR]{BKKR10}
forced turbulence have shown that the
effective magnetic pressure is negative for field strengths below about
40\% of the equipartition value, provided the magnetic Reynolds number
exceeds unity.
However, definitive proof of an instability associated with the negative
effective magnetic pressure phenomenon came only more recently with DNS
of forced turbulence that have sufficiently many turbulent eddies
in the simulation domain \citep{BKKMR11,KBKMR12}.
This work has only become possible due to earlier DNS \citep[BKR,][]{KBKR11}
exploring first the relevant parameter regime.
In the present study we
investigate the effect of turbulent convection on the
effective mean Lorentz force in DNS and study the formation of large-scale
magnetic structures in mean-field models.

\section{Effective mean Lorentz force}

In this section we state the underlying equations, highlighting the
difference to earlier work where anisotropic contributions from gravity
were either weak or absent.

\subsection{Governing equations}

In this study we are mainly interested in the effects
of turbulent convection on the mean Lorentz force.
To this end we consider the momentum equation,
\EQ
{\partial\over\partial t}\rho\, U_i=-{\partial\over
\partial x_j}\Pi_{ij} + \rho \, g_i,
\EN
where ${\bm g}$ is the acceleration due to gravity,
\EQ
\Pi_{ij}=\rho \, U_iU_j+\delta_{ij}\left(p
+\half\BB^2\right)-B_iB_j-2\nu \rho \, {\sf S}_{ij},
\label{Piorig}
\EN
is the momentum stress tensor, $\UU$ and $\BB$ are
the velocity and
magnetic fields, $p$ and $\rho$ are the fluid pressure
and density,
$\delta_{ij}$ is the Kronecker tensor, $\nu$ is the
kinematic viscosity, and
\EQ
{\sf S}_{ij}=\half(\partial_i U_j+\partial_j U_i)-\onethird\delta_{ij}
\nab\cdot\UU
\label{strain}
\EN
is the trace-free rate of strain tensor.
Throughout this paper we have adopted units where the
vacuum permeability $\mu_0$ is set to unity, although we
do include it in some expressions for clarity.

Neglecting correlations between velocity and density fluctuations
for low-Mach number turbulence, the averaged momentum
equation is
\EQ
{\partial\over\partial t} \meanrho \, \meanU_i =
-{\partial\over\partial x_j}\overline{\Pi}_{ij}
+ \meanrho \, g_i,
\EN
where
$\meanrho$ is the mean fluid density,
$\meanUU$ is the mean fluid velocity,
$\overline\Pi_{ij}=\overline\Pi_{ij}^{\rm m}
+\overline\Pi_{ij}^{\rm f}$ is the mean momentum stress
tensor split into contributions resulting
entirely from the mean field (indicated by superscript m)
and those of
the fluctuating field (indicated by superscript f).
The tensor $\overline\Pi_{ij}^{\rm m}$ has the same form as \Eq{Piorig},
but all quantities have now attained an overbar, i.e.\
\EQ
\overline\Pi_{ij}^{\rm m}=\meanrho \, \meanU_i\meanU_j
+\delta_{ij}\left(\overline{p}+\half\meanBB^2\right)
-\meanB_i\meanB_j-2\nu \meanrho \, \overline{\sf S}_{ij},
\EN
where
$\meanBB$ is the mean magnetic field
and $\overline{p}$ is the mean fluid pressure.
The contributions, $\overline\Pi_{ij}^{\rm f}$, which
result from the fluctuations of velocity $\uu=\UU-\meanUU$ and
magnetic fields $\bb=\BB-\meanBB$, are determined by
\EQ
\overline\Pi_{ij}^{\rm f}=\meanrho \, \overline{u_iu_j}
+\half\delta_{ij}\overline{\bb^2}
-\overline{b_ib_j}.
\label{stress0}
\EN
This contribution, together with the one
from the mean field,
$\overline\Pi_{ij}^{\rm m}$, comprises the total
mean momentum tensor.
The contribution from the fluctuating fields is split into
parts that are independent of the mean magnetic
field (which determine the turbulent viscosity and
background turbulent pressure)
and parts that do depend on the mean magnetic field.

In the present study we consider turbulent convection with
an imposed uniform horizontal magnetic field, $\BB_0=(B_0,0,0)$,
that is perpendicular to the direction of gravity.
This modifies the stress tensor from $\overline\Pi_{ij}^{\rm f,0}$
to $\overline\Pi_{ij}^{{\rm f},\overline{B}}$, so only the difference,
\begin{eqnarray}
\Delta\overline\Pi_{ij}^{{\rm f}}\equiv
\overline\Pi_{ij}^{{\rm f},\overline{B}}
-\overline\Pi_{ij}^{\rm f,0},
\label{stress-par2}
\end{eqnarray}
depends on the mean magnetic field $\meanBB$ and can be parameterized
as \citep{RK07}
\begin{eqnarray}
\Delta\overline\Pi_{ij}^{{\rm f}}=
 \qs\meanBB^2 \Beta_i\Beta_j
- \, \Big(\half \qp \, \delta_{ij} + \qg \,
\hat g_i \hat g_j \Big) \meanBB^2,
\label{stress-par}
\end{eqnarray}
where $\hat {\bm g}$ is the vertical unit vector directed
along the gravity field,
$\Beta_j = \meanB_j /\meanB$
is the unit vector
directed along the mean magnetic field, $\qs$, $\qp$,
and $\qg$ are functions of magnetic Reynolds and Prandtl numbers
as well as the modulus of the normalized mean field,
\EQ
\beta=\meanB/\Beq,\quad\mbox{where}\quad
\meanB=|\meanBB|,\quad\mbox{and}\quad
B_{\rm eq}=(\overline{\rho\uu^2})^{1/2}
\EN
is the equipartition field strength.
To derive Eq.~(\ref{stress-par}),
we use Eqs.~(A22)--(A24) of \cite{RK07}.
The parametrization~(\ref{stress-par}) follows also from
symmetry arguments which allow us to construct
a symmetric tensor with two preferential perpendicular
directions along the horizontal magnetic field $\BBeta$
and vertical gravity field $\hat {\bm g}$.
Such symmetric tensor is a linear combination of symmetric
tensors $\delta_{ij}$, $\Beta_i\Beta_j$ and
$\hat g_i \hat g_j$.
(In the case of an oblique imposed magnetic
field, there would be an additional contribution.)

The effective mean Lorentz force that takes into
account the turbulent convection effects, reads:
\begin{eqnarray}
\meanrho \, \meanFFF^{\rm M}_i &=& -\nabla_j
\Big(\half\meanBB^2 \delta_{ij} -\meanB_i\meanB_j
+\Delta\overline\Pi_{ij}^{\rm f}\Big)
\nonumber\\
&=& -\half\nabla_i \Big[(1-\qp) \, \meanBB^2\Big]
+ \hat g_i \, \nabla_z \Big(\qg \,  \meanBB^2\Big)
\nonumber\\
&& + \meanBB \cdot \bm\nabla \Big[(1-\qs)
\,\meanBB\Big].
\label{Lor-force}
\end{eqnarray}
The analytic expressions for the nonlinear
quenching functions, $\qp(\beta)$, $\qs(\beta)$, and $\qg(\beta)$
for turbulent convection have been derived in
\cite{RK07}.
Their asymptotic formulae are given below.
For weak mean magnetic fields, $4\beta \ll \Rm^{-1/4}$, the
functions $\qp$, $\qs$, and $\qg$ are given by
\begin{eqnarray*}
\qp(\beta) &=& {4 \over 5} \,  \biggl(\ln \Rm +
{4 \over 45} \biggr) - {7  \over 3} \, a_\ast
+ {16 \, \ell_0^2 \over 9 \, H_\rho^2},
\\
\qs(\beta) &=& {8 \over 15} \,  \biggl(\ln \Rm +
{2 \over 15} \biggr) ,
\quad \qg(\beta) = 8 a_\ast - {8 \, \ell_0^2 \over 3 \, H_\rho^2};
\end{eqnarray*}
for $\Rm^{-1/4} \ll 4\beta \ll 1$ these functions are
\begin{eqnarray*}
\qp(\beta) &=& {16 \over 25} \,
\Big[5|\ln (4 \beta)|+ 1 + 32 \, \beta^{2}\Big]
- {7  \over 3} \, a_\ast + {16 \, \ell_0^2 \over 9
\, H_\rho^2},
\\
\qs(\beta) &=& {32 \over 15} \, \biggl[|\ln (4 \beta)|
+{1 \over 30} + 12  \beta^{2} \biggr],
\\
\qg(\beta) &=& 8 a_\ast - {8 \, \ell_0^2 \over 3
\, H_\rho^2},
\end{eqnarray*}
while for strong fields, $4\beta \gg 1 $, they are
\begin{eqnarray*}
\qp(\beta) &=& {1 \over 6\beta^2} \, \Big(1
+ {3 \, \ell_0^2 \over \, H_\rho^2} \Big) +
{\pi a_\ast \over 80 \beta} ,
\\
\qs(\beta) &=& {\pi \over 48  \beta^3}
+ {3 \pi a_\ast \over 160  \beta} ,
\quad
\qg(\beta) = {3 \pi a_\ast \over 40  \beta^3}
- {3 \, \ell_0^2 \over 4 \, H_\rho^2 \, \beta^2}.
\end{eqnarray*}
Here, $\Rm=\ell_0 \, u_{\rm ums} / \eta$ is the magnetic
Reynolds number based on the integral scale of turbulent
convection, $\ell_0$, and the root-mean-square (rms) value
of the velocity, $\urms$, $\,\eta$ is the magnetic diffusion
due to the electrical conductivity of the fluid,
and $H_{\rho}$ is the density scale height.
The parameter $a_\ast$ characterizes turbulent convection and is
determined from the budget equation for the total energy, yielding
\begin{eqnarray*}
a_\ast^{-1} = 1 + \left[\nu_{\rm t} (\nab \meanUU)^2
+ \eta_{\rm t} (\nab \meanBB)^2 /\rho\right]/ g F_\ast \;,
\end{eqnarray*}
where $\nu_{\rm t}$ is the turbulent viscosity, $\eta_{\rm t}$
is the turbulent magnetic diffusivity,
$F_\ast = \overline{u_z \, s'}$ is the
vertical heat flux from the background turbulent convection,
and $s'$ are fluctuations of the specific entropy.

\subsection{Turbulent contributions to effective Lorentz force}
\label{TurbSim}

To study the effects of turbulent convection on the
Reynolds and Maxwell stresses, and hence on the effective
Lorentz force from the mean field, we need to determine the functions
$\qp(\beta)$, \, $\qs(\beta)$, and $\qe(\beta)$ in DNS.
Allowing here for the possibility of small-scale dynamo action,
Eqs.~(\ref{stress0})--(\ref{stress-par}) yield
\begin{eqnarray}
&& \Delta\overline{\rho u_iu_j}
 + \half\delta_{ij}\Delta\overline{\bb^2}
-\Delta\overline{b_ib_j}
\nonumber\\
&& = \left[\qs \Beta_i\Beta_j
- \left(\half \qp \, \delta_{ij} + \qg \, \hat g_i \,
\hat g_j \right)\right] \meanBB^2.
\label{funct-tenzor}
\end{eqnarray}
Here, $\Delta\overline{\rho u_iu_j}=\overline{\rho u_iu_j}
-\overline{\rho u_{0i}u_{0j}}$ and
$\Delta\overline{b_ib_j}=\overline{b_ib_j}-\overline{b_{0i}b_{0j}}$,
where subscripts 0 indicate
values in the absence of the mean magnetic field.
To obtain 3 independent equations for the 3 unknowns,
we multiply Eq.~(\ref{funct-tenzor}):\\
(i) by $\hat g_i \, \hat g_j$,
\begin{equation}
\Delta\overline{\rho u_z^2}
+ \half\Delta\overline{\bb^2} -\Delta\overline{b_z^2}
= - \Big(\half \qp + \qg\Big) \meanBB^2 ,
\label{funct-z}
\end{equation}
(ii) by $\Beta_i\Beta_j$ (defining $u_{\hatBB}=\uu\cdot\hatBB_0$
and $b_{\hatBB}=\bb\cdot\hatBB_0$):
\begin{equation}
\Delta\overline{\rho u_{\hatBB}^2}
+ \half\Delta\overline{\bb^2} -\Delta\overline{b_{\hatBB}^2} = -
\Big(\half\qp - \qs\Big) \meanBB^2,
\label{funct-bet}
\end{equation}
(iii) and compute the trace of Eq.~(\ref{funct-tenzor}),
\begin{equation}
\Delta\overline{\rho\uu^2}
+ \half\Delta\overline{\bb^2}
= - \Big( {\textstyle{3 \over 2}} \qp + \qg - \qs\Big) \meanBB^2.
\label{funct-trace}
\end{equation}
Equations~(\ref{funct-z})--(\ref{funct-trace}) yield the functions
$\qp(\beta)$, $\qs(\beta)$, and $\qg(\beta)$:
\EQ
\qp(\beta) \, \meanB^2 = -2 \Delta\overline{\rho u_y^2}
- \Delta\overline{\bb^2} + 2 \Delta\overline{b_y^2},
\label{stress3}
\EN
\EQ
\qs(\beta) \, \meanB^2 = - \Delta\overline{\rho u_y^2}
+ \Delta\overline{\rho u_x^2}
+ \Delta\overline{b_y^2}
-\Delta\overline{b_x^2},
\label{stress4}
\EN
\EQ
\qg(\beta) \, \meanB^2 = - \Delta\overline{\rho u_z^2}
+ \Delta\overline{\rho u_y^2}
+ \Delta\overline{b_z^2}
- \Delta\overline{b_y^2}.
\label{stress5}
\EN
Using Eqs.~(\ref{stress3})--(\ref{stress5}), we
determine the functions $\qp(\beta)$, $\qs(\beta)$,
and $\qg(\beta)$ from DNS in \Sec{DNS}.
In all those cases, overbars denote averages over $x$, $y$, and $t$
during the statistically steady state, corresponding to a
time interval $\Delta t$ of up to a thousand turnover times, i.e., $\Delta t\urms\kf=O(1000)$.

\section{Direct numerical simulations}
\label{DNS}
\subsection{DNS model}
\begin{figure*}
\begin{center}
\includegraphics[width=\textwidth]{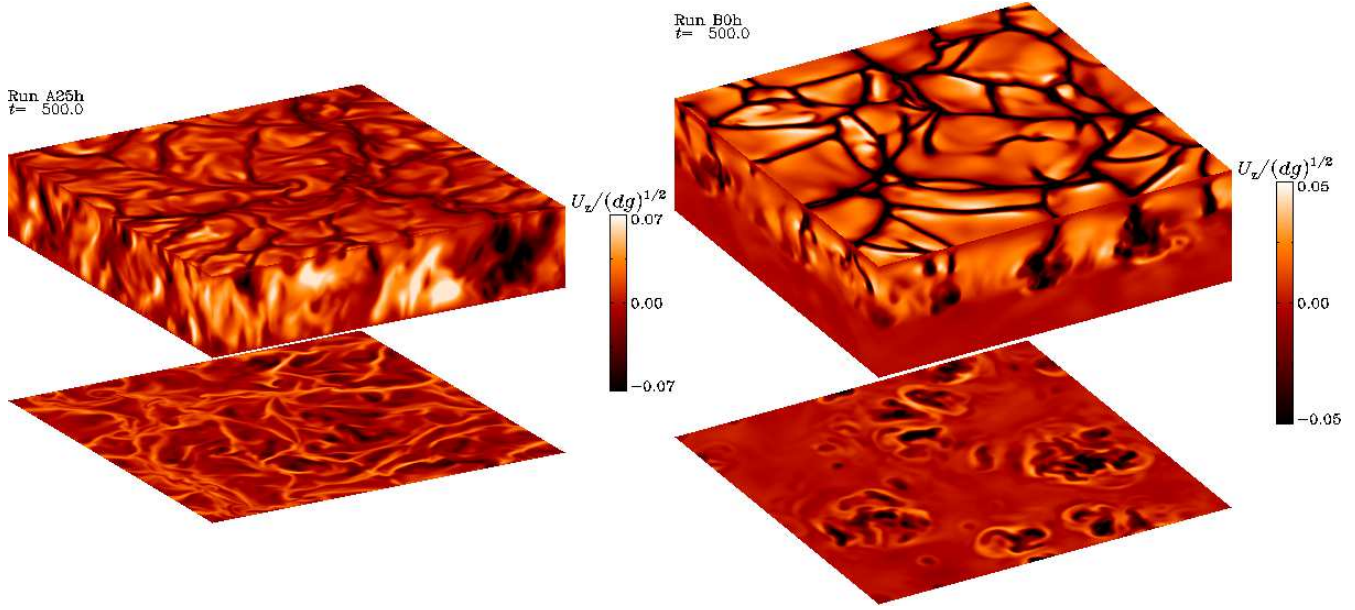}
\end{center}
\caption[]{Velocity component $U_z$ at the periphery of the domain
  from hydrodynamical runs A25h ($\Delta\rho=1.2$, left)
  and B0h ($\Delta\rho=320$, right). In both cases
  the top and bottom slices show $U_z$ near the top and bottom of the
  convectively unstable layer, respectively.}
\label{boxes}
\end{figure*}

We use two different setups when studying the effects of convection on
the mean Lorentz force: (i) a weakly stratified model, similar
to that used in \cite{KKB10}, without overshoot layers, and (ii)
a strongly stratified setup, similar to that in \cite{KKB09a}, including
upper and lower overshoot layers. In both cases we use a Cartesian
domain with $L_x=L_y=5d$, where $d$ is the depth of the convectively
unstable layer. The convective layer is situated between $0<z<d$ in
both setups. In the strongly stratified setup the $z$ coordinate runs
from $-0.85d<z<1.15$.
We solve
a set of hydromagnetic equations
\begin{eqnarray}
\frac{\pd \bm{A}}{\pd t} &\!=\!& \bm{U}\times\bm{B} - \eta \mu_0\bm{J}, \\
\frac{\DD \bm{U}}{\DD t} &\!=\!& -\frac{1}{\rho}{\bm \nabla}p
+ {\bm g} + \frac{1}{\rho} \bm{J} \times {\bm B}
+ \frac{1}{\rho} \bm{\nabla} \cdot 2 \nu \rho
\mbox{\boldmath ${\sf S}$} , \label{equ:mom}\\
\frac{\DD \ln \rho}{\DD t} &\!=\!& -\bm\nabla\cdot\bm{U}, \\
T\frac{\DD s}{\DD t} &\!=\!&\frac{1}{\rho} \bm{\nabla} \cdot K \bm{\nabla}T + 2 \nu
\mbox{\boldmath ${\sf S}$}^2 + \frac{\eta \mu_0}{\rho} \bm{J}^2 - \Gamma_{\rm cool},
\end{eqnarray}
where $\DD/\DD t=\pd/\pd t + \bm{U}\cdot\bm\nabla$ is the
advective time derivative, $\bm{A}$ is the magnetic
vector potential, $\bm{B} =
\bm{\nabla} \times \bm{A} + \bm{B}_0$ the magnetic field,
$\bm{B}_0=(B_0,0,0)$ is the imposed field, $\bm{J} =
\mu_0^{-1} \bm{\nabla} \times \bm{B}$ is the current density,
$\eta$ and $\nu$ are the magnetic
diffusivity and kinematic viscosity, respectively,
$K$ is the heat
conductivity, $\rho$ is the density, $\bm{U}$
is the velocity, $s$ is the specific entropy, and
$\bm{g} = -g\hat{\bm{z}}$ is the gravitational acceleration.
The fluid obeys an ideal gas law $p=\rho e (\gamma-1)$,
where $p$ and $e$ are
the pressure and internal energy, respectively, and
$\gamma = c_{\rm P}/c_{\rm V} = 5/3$ is the ratio of specific
heats at constant
pressure and volume, respectively.
The specific internal energy per
unit mass is related to the temperature via $e=c_{\rm V} T$,
and the rate of strain tensor $\mbox{\boldmath ${\sf S}$}$
is given by
Eq.~(\ref{strain}).
The stratification in the
hydrostatic initial
state can be described by a polytrope with index $m=1$ in the weakly
stratified case, and stratified three-layer setup is described by
polytropic indices $(m_1,m_2,m_3)=(3,1,1)$.
In the latter setup, a cooling term
$\Gamma_{\rm cool}$ operates in the region $z>1$,
keeping the layer isothermal.
The density contrast across the full domain,
$\Delta\rho=\rho_{\rm bot}/\rho_{\rm top}$, is
1.2 (runs A) and 320 (runs B, see Table~\ref{runs}).
In the latter case, the density changes by a factor of roughly
13 within the convectively unstable layer.
Typical flow patterns from both setups are shown in \Fig{boxes}.
In all simulations we use the {\sc Pencil
Code}\footnote{http://pencil-code.googlecode.com}.

\begin{table}
\caption{Summary of the runs.
Here, $\Ma = \urms/\sqrt{dg}$, and the imposed field in
normalized form is given by $\tilde{B}_0=B_0/\Beq$,
where $\Beq$ is the volume averaged
equipartition field. The last column gives the density
contrast, $\Delta \rho$, over the entire domain.
The Prandtl number is equal to unity in all runs.
We use grid resolutions $128^2\times64$ (Runs~A1--A24),
$256^2\times128$ (Runs~A25--A29), and
$256^2\times192$ (Runs~B1--B8).
}
\vspace{12pt}
\centerline{\begin{tabular}{lccccccc}
    Run & $\Ma$ & ${\rm Ra}$ & $\Rem$ & $\Pm$ & $\tilde{B}_0$ & $\Delta \rho$ \\
    \hline
    A0h   & 0.068 & $10^6$ &--- & --- &  --- &  1.2   \\
    \hline
    A1    & 0.068 & $10^6$ & 11 & 0.2 & 0.07 &  1.2   \\ 
    A2    & 0.066 & $10^6$ & 11 & 0.2 & 0.15 &  1.2   \\ 
    A3    & 0.059 & $10^6$ &  9 & 0.2 & 0.34 &  1.2   \\ 
    A4    & 0.055 & $10^6$ &  9 & 0.2 & 0.54 &  1.2   \\ 
    A5    & 0.052 & $10^6$ &  8 & 0.2 & 0.78 &  1.2   \\ 
    A6    & 0.053 & $10^6$ &  8 & 0.2 & 0.95 &  1.2   \\ 
    A7    & 0.059 & $10^6$ &  9 & 0.2 & 1.17 &  1.2   \\ 
    A8    & 0.066 & $10^6$ & 11 & 0.2 & 1.52 &  1.2   \\ 
    \hline
    A9    & 0.068 & $10^6$ & 22 & 0.4 & 0.07 &  1.2   \\ 
    A10   & 0.065 & $10^6$ & 21 & 0.4 & 0.15 &  1.2   \\ 
    A11   & 0.055 & $10^6$ & 18 & 0.4 & 0.36 &  1.2   \\ 
    A12   & 0.051 & $10^6$ & 16 & 0.4 & 0.59 &  1.2   \\ 
    A13   & 0.048 & $10^6$ & 15 & 0.4 & 0.83 &  1.2   \\ 
    A14   & 0.059 & $10^6$ & 16 & 0.4 & 1.01 &  1.2   \\ 
    A15   & 0.061 & $10^6$ & 19 & 0.4 & 1.15 &  1.2   \\ 
    A16   & 0.066 & $10^6$ & 21 & 0.4 & 1.51 &  1.2   \\ 
    \hline
    A17   & 0.065 & $10^6$ & 51 & 1.0 & 0.08 &  1.2   \\ 
    A18   & 0.061 & $10^6$ & 49 & 1.0 & 0.16 &  1.2   \\ 
    A19   & 0.051 & $10^6$ & 41 & 1.0 & 0.39 &  1.2   \\ 
    A20   & 0.047 & $10^6$ & 37 & 1.0 & 0.64 &  1.2   \\ 
    A21   & 0.047 & $10^6$ & 37 & 1.0 & 0.86 &  1.2   \\ 
    A22   & 0.050 & $10^6$ & 39 & 1.0 & 1.01 &  1.2   \\ 
    A23   & 0.059 & $10^6$ & 47 & 1.0 & 1.18 &  1.2   \\ 
    A24   & 0.069 & $10^6$ & 55 & 1.0 & 1.46 &  1.2   \\ 
    \hline
    A25h  & 0.058 & $4.2\cdot10^6$ &  - &  -  &  -   &  1.2   \\ 
    \hline
    A25   & 0.058 & $4.2\cdot10^6$ & 92 & 1.0 & 0.09 &  1.2   \\ 
    A26   & 0.044 & $4.2\cdot10^6$ & 70 & 1.0 & 0.46 &  1.2   \\ 
    A27   & 0.040 & $4.2\cdot10^6$ & 63 & 1.0 & 0.76 &  1.2   \\ 
    A28   & 0.042 & $4.2\cdot10^6$ & 66 & 1.0 & 0.96 &  1.2   \\ 
    A29   & 0.047 & $4.2\cdot10^6$ & 75 & 1.0 & 1.07 &  1.2   \\ 
    \hline
    B0h   & 0.032 & $1.2\cdot10^7$ & -  &  -  &  -   &  320   \\ 
    \hline
    B1    & 0.032 & $1.2\cdot10^7$ & 51 & 1.0 & 0.01 &  320   \\ 
    B2    & 0.031 & $1.2\cdot10^7$ & 49 & 1.0 & 0.07 &  320   \\ 
    B3    & 0.028 & $1.2\cdot10^7$ & 45 & 1.0 & 0.23 &  320   \\ 
    B4    & 0.027 & $1.2\cdot10^7$ & 43 & 1.0 & 0.32 &  320   \\ 
    B5    & 0.027 & $1.2\cdot10^7$ & 43 & 1.0 & 0.41 &  320   \\ 
    B6    & 0.026 & $1.2\cdot10^7$ & 42 & 1.0 & 0.61 &  320   \\ 
    B7    & 0.025 & $1.2\cdot10^7$ & 40 & 1.0 & 0.78 &  320   \\ 
    B8    & 0.025 & $1.2\cdot10^7$ & 40 & 1.0 & 0.92 &  320   \\ 
    \hline
\label{runs}\end{tabular}}\end{table}

The horizontal boundaries are periodic.
In the weakly stratified case we keep the temperature fixed at the top
and bottom boundaries, whereas in the more strongly stratified setup the upper
boundary is isothermal and a constant flux of energy is applied at the
lower boundary by fixing the temperature gradient. For the velocity we
apply impenetrable, stress-free conditions according to
\begin{eqnarray}
\pd_zU_x = \pd_z U_y = U_z=0.
\end{eqnarray}
For the magnetic field we use vertical field conditions
\begin{eqnarray}
B_x-B_0 = B_y=0.
\end{eqnarray}
Dimensionless quantities are obtained by setting
\begin{eqnarray}
d = g = \rho_0 = c_{\rm P} = \mu_0 = 1\;,
\end{eqnarray}
where $\rho_0$ is the fluid density at $z_{\rm m}=\half d$.
The units of length, time, velocity, density, specific
entropy, and magnetic field are then
\begin{eqnarray}
&& [x] = d\;,\;\; [t] = \sqrt{d/g}\;,\;\; [U]=\sqrt{dg}
\;,\;\;
\nonumber \\
&& [\rho]=\rho_0\;,\;\; [s]=c_{\rm P}\;,\;\; [B]=\sqrt{dg\rho_0\mu_0}\;.
\end{eqnarray}
The simulations are controlled by the following
dimensionless parameters: thermal and magnetic diffusion in
comparison to viscosity are measured by the Prandtl numbers
\begin{eqnarray}
\Pr=\frac{\nu}{\chi_0}, \quad \Pm=\frac{\nu}{\eta},
\end{eqnarray}
where $\chi_0=K/(c_{\rm P} \rho_0)$ is the reference value
of the thermal diffusion coefficient measured in the
middle of the layer
($z_{\rm m}$) of the non-convecting hydrostatic reference initial state.
The efficiency of convection is characterized by the
Rayleigh number
\begin{eqnarray}
\Ra=\frac{g d^4}{\nu \chi_0}\left(- \frac{1}{c_{\rm P}}
\frac{{\rm d}s}{{\rm d}z} \right)_{z_{\rm m}},
\end{eqnarray}
which is again determined from the initial non-convecting state at
$z_{\rm m}$.
The entropy gradient can be presented in terms of
logarithmic temperature gradients
\begin{eqnarray}
\left(- \frac{1}{c_{\rm P}}\frac{{\rm d}s}{{\rm d}z}
\right)_{z_{\rm m}}=\frac{\nabla-\nabla_{\rm ad}}{H_{\rm P}},
\end{eqnarray}
with $\nabla=(\pd \ln T/\pd \ln p)_{z_{\rm m}}$, $\nabla_{\rm ad}
=1-1/\gamma$, and $H_{\rm P}$ being the pressure scale
height at $z=z_{\rm m}$.

The effects of viscosity and magnetic
diffusion are quantified respectively by the fluid and
magnetic Reynolds numbers
\begin{eqnarray}
\Rey=\frac{\urms}{\nu \kf}, \quad \Rem=\frac{\urms}{\eta \kf}=\Pm\,\Rey,
\end{eqnarray}
where $\urms$ is the root-mean-square (rms) value of the
velocity and $\kf=2\pi/d$ is the wavenumber corresponding
to the depth of the convectively unstable layer.
Again, it is convenient to measure the magnetic field strength
in terms of the equipartition value.
The values of these parameters used in different runs
are given in Table~\ref{runs}.

\subsection{Effective mean Lorentz force from DNS}

We now turn to DNS models of turbulent convection
to determine the coefficients $\qp$, $\qs$,
and $\qg$ using Eqs.~(\ref{stress3})--(\ref{stress5}).
First, we perform purely hydrodynamical simulations
to determine the turbulent background velocity
$\overline{\uu^2}$ in the absence of magnetic fields.
No dynamo action occurs in Runs~A1--A16 and B1--B8, whereas in
Runs~A17--A24 the dynamo is growing very slowly, and in A25--A29 a
small-scale dynamo is operating. We find that the critical $\Rem$ for
$\Pm=1$ is between 50 and 60, which is almost two times higher than in
the case $\Pm=5$ studied earlier \citep{KKB08}.
The hydrodynamical simulations have been run sufficiently long
(roughly 250 turnover times) so that it is thermally relaxed
and the turbulence is statistically steady.
The last snapshot of the hydrodynamical run
is used as the initial condition for all subsequent simulations
where a uniform magnetic field is imposed.

Next, we consider an imposed horizontal field,
$\BB_0=(B_0,0,0)$.
Earlier numerical studies have shown that
the effective mean magnetic pressure that takes into
account turbulence effects,
is negative when the mean magnetic fields
are smaller than
the equipartition strength in non-stratified
(BKR) and stratified (BKKR) forced turbulence.
In the present study we investigate this issue
for turbulent convection
with weak and strong density stratification.
Let us define the dimensionless effective
mean magnetic pressure $\Peff$ as
\begin{equation}
\Peff = \half(1-\qp)\meanv{B}^2\!/\Beq^2,
\end{equation}
where $\qp=\qp(\mean{B}/\Beq)$ and $\Beq=\Beq(z)$.
Following earlier work \citep[BKKR,][]{KBKMR12}, we characterize our
numerical results for $\qp$ by a fit of the form
\EQ
\qp=\frac{\qpz}{1+\BB^2\!/B_{\rm p}^2},
\label{fit}
\EN
where $\qpz$ and $\Bp$ are fit parameters,
which are determined by matching the shape of $\Peff$ near its minimum
and where $\dd\Peff/\dd\meanB^2<0$, which is the range relevant to the
negative magnetic pressure instability \cite[][BKKR]{RK07}.
We use the same ansatz also for $\qs$ and $\qe$, and define
in this way the fit parameters $\qsz$, $\qgz$, $\Bs$, and $\Bg$.
Note that, due to turbulent pumping effects, the mean
magnetic field $\meanv{B}$ in general also depends
on height---even in the absence of a large-scale
dynamo; see \cite{KKB10} and BKKR.

\begin{figure}
\begin{center}
\includegraphics[width=\columnwidth]{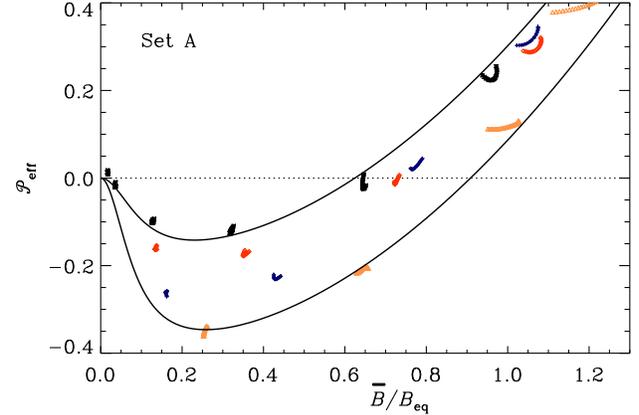}
\end{center}
\caption[]{Effective magnetic pressure as a function of
the mean magnetic field from weakly stratified Runs~A1--A29
with an imposed horizontal field $\BB_0=B_0\hat{\bm{x}}$.
The black stars, red diamonds, blue crosses, and yellow triangles
denote simulations with $\Rem\approx10$, 20, 50, and 70,
respectively.
We omit points near the boundaries at $z/d<0.35$ and $z/d>0.65$.
The two curves correspond to approximate fits
determined by Eq.~(\ref{fit}), with $\qpz=40$ and $\Bp=0.1\Beq$
(upper curve, small $\Rem$), and $\qpz=130$ and $\Bp=0.08\Beq$
(lower curve, larger $\Rem$), respectively.}
\label{ppAA1}
\end{figure}

Results for the effective mean magnetic pressure
$\Peff$ from weakly and strongly
stratified runs are shown in \Figss{ppAA1}{ppAAA}.
Since $\meanv{B}$ and $\Beq$ are functions of $z$, we obtain
for each combination of imposed field strength and $\Rem$
a family of solutions for $\qp$, $\qs$, and $\qe$.
We neglect points near the top and bottom boundaries
to avoid boundary effects.
The result is shown in \Fig{ppAA1}.
For weak stratification (Runs~A1--A29), a negative contribution of
turbulent convection to the mean magnetic pressure is found if the
mean magnetic field is smaller than the equipartition value.
The maximum of this contribution is attained near
$\meanv{B}\approx0.5\Beq \hat{\bm{x}}$ and it tends to
be somewhat stronger for larger magnetic Reynolds number.

\begin{figure}
\begin{center}
\includegraphics[width=\columnwidth]{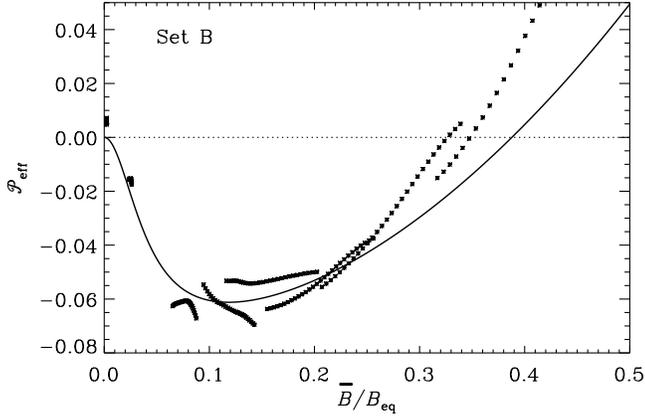}
\end{center}
\caption[]{Same as Fig.~\ref{ppAA1} but for Runs~B1--B8 for $\Rem=40$--$50$.
The solid line corresponds to a fit with $\qpz=95$ and $\Bp=0.04\Beq$.
}\label{ppAAA}
\end{figure}

In the strongly stratified runs (see \Fig{ppAAA})
we also find a negative contribution
of turbulent convection to the mean magnetic pressure,
but it is constrained to somewhat lower values
($\meanv{B}<0.4\Beq \hat{\bm{x}}$) of
the mean magnetic field, and the effective mean magnetic
pressure, $\Peff$, has a weaker minimum
than in the weakly stratified case. It appears
that we find universal scaling
for $\Peff$ as a function of $\meanv{B}/\Beq$
as was obtained in BKKR for stratified forced turbulence,
provided that only data points near the middle ($0.35<z/d<0.65$) of the
convectively unstable layer are used.
Furthermore, our highly stratified simulations show that
the minimum of $\Peff$ and the range
of the mean magnetic field in which $\Peff$
is negative, are roughly consistent with
those found by BKKR.

\begin{figure}
\begin{center}
\includegraphics[width=\columnwidth]{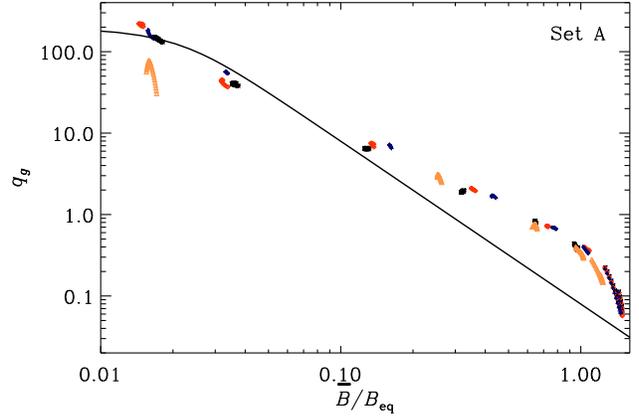}
\end{center}
\caption[]{$\qe$ as a function of mean
  magnetic field from Runs~A1--A29. The symbols and colours
  are the same as in Fig.~\ref{ppAA1}.
  The solid line applies to the fit parameters $\qgz=200$ and $\Bg=0.025\Beq$.}
\label{pqeAA1}
\end{figure}

Due to anisotropy of turbulent convection there is
an significant contribution to the effective mean
magnetic pressure characterized by the term $\qe$.
This function affects the vertical
component of the effective mean Lorentz force;
see the second term on the right-hand-side
of Eq.~(\ref{Lor-force}).
Our DNS show that the dimensionless quantity $\qe$
is mostly positive (see \Figs{pqeAA1}{pqeAAA}), which
implies that this effect increases the negative
contribution of anisotropic turbulent convection
to the effective mean magnetic pressure.
Note that the DNS in stratified forced turbulence
of BKKR has not found strong anisotropic contributions
as characterized by the term $\qe$.
On the other hand, the negative
contribution of anisotropic turbulent convection
to the effective mean magnetic tension, as characterized
by positive values of $\qs$, has neither been found in our DNS
(see \Figs{pteAA1}{pteAAA}) nor in those of BKKR.

\begin{figure}
\begin{center}
\includegraphics[width=\columnwidth]{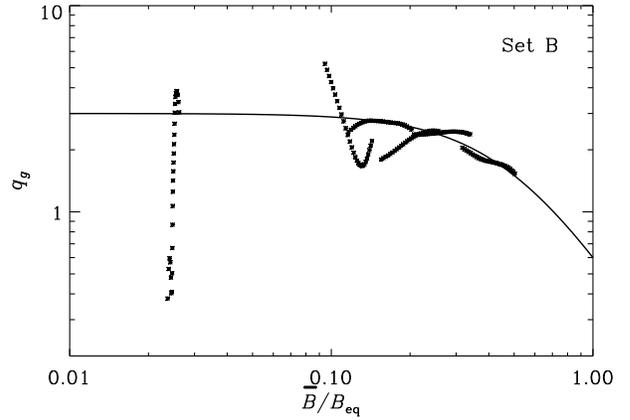}
\end{center}
\caption[]{Same as Fig.~\ref{pqeAA1} but for Runs~B1--B8.
The fit parameters are $\qez=3$ and $\Bg=0.5\Beq$.
}\label{pqeAAA}
\end{figure}

\begin{figure}
\begin{center}
\includegraphics[width=\columnwidth]{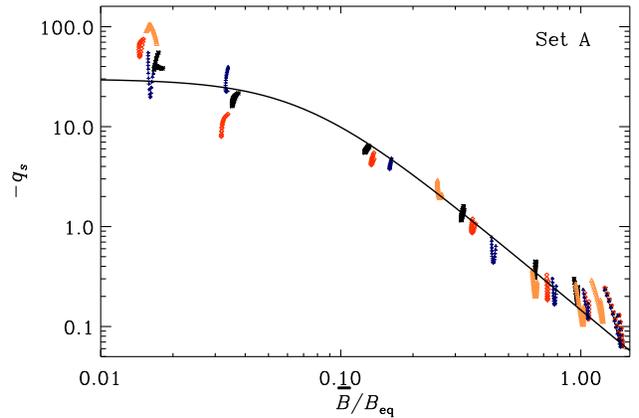}
\end{center}
\caption[]{Effective magnetic tension parameter as a function of the mean
  magnetic field from Runs~A1--A29. The symbols and colours
  are the same as in Fig.~\ref{ppAA1}.
  The solid line applies to the fit parameters $\qsz=-30$ and $\Bs=0.07\Beq$.}
\label{pteAA1}
\end{figure}

\begin{figure}
\begin{center}
\includegraphics[width=\columnwidth]{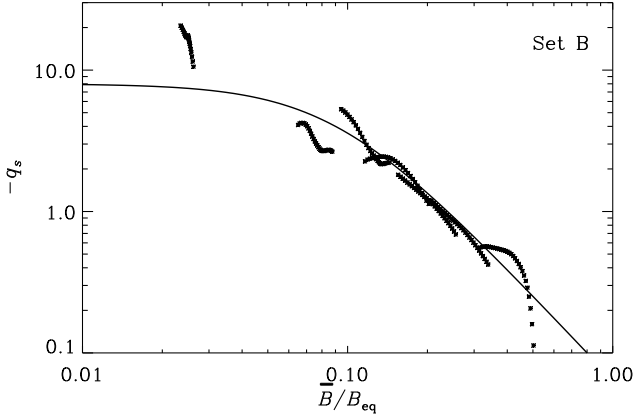}
\end{center}
\caption[]{Same as Fig.~\ref{pteAA1} but for Runs~B1--B8.
The solid line applies to the fit parameters $\qsz=-8$ and $\Bs=0.09\Beq$.}
\label{pteAAA}
\end{figure}

\begin{figure}
\begin{center}
\includegraphics[width=0.55\columnwidth,angle=-90]{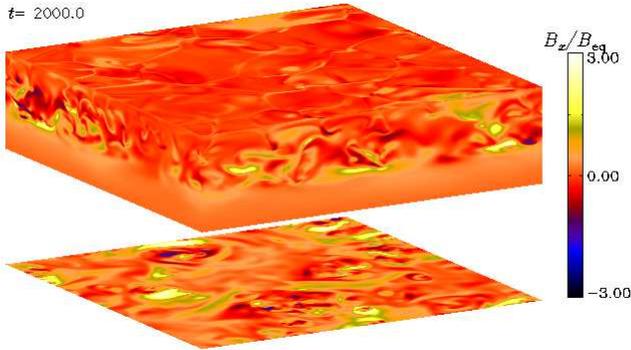}
\end{center}
\caption[]{Magnetic field component $B_x$ from Run~B3
with imposed horizontal field $B_0\hat{\bm{x}}$
at $\Rem=45$, $\Pm=1$, a density contrast of 320, and $B_0/\Beq=0.23$.}
\label{o256x192b3_Bx}
\end{figure}

In \FFig{o256x192b3_Bx} we show the magnetic field component
$B_x$ from Run~B3 with an imposed horizontal
magnetic field $\meanv{B}\approx 0.23\Beq\hat{\bm{x}}$.
The structure of the magnetic field, however, does not
show clear signs of magnetic flux concentrations in the DNS.
Even after additional averaging over time and along the $x$ direction
no spatial modulation of the magnetic field is seen.
The simulations of \cite{BKKMR11} strongly suggest that
the reason for this is related to lack of scale separation.
As demonstrated in Fig.~17 of BKKR, at larger scale separation the
turbulent diffusivity on the scale of the domain becomes weak enough
to allow for the development of large-scale magnetic structures.

\section{Entropy evolution in mean-field models}

\subsection{Mean-field equations}

We now apply a mean-field model similar to that of BKR for
adiabatic stratification and those of BKKR and \cite{KBKR11}
for isothermal stratification.
Both types of models are in principle able to display a large-scale
instability provided the domain is big compared with the typical size
of turbulent eddies, i.e., the scale separation ratio is large.
In the mean-field calculations of BKKR it was shown that in models with
too low scale separation ratio the turbulent magnetic diffusivity
and turbulent viscosity (which are proportional to the scale of the
energy-carrying turbulent eddies) was too large, so the instability
is too weak or not excited.
Even if the scale separation ratio is big enough, the instability
can only develop if $\dd\Peff/\dd\beta^2$, taken at the value of the
imposed field, is negative inside the domain \cite[][BKKR]{KBKR11}.
If these conditions are satisfied, the maximum growth rate of the
instability was shown to be independent of the strength of the
imposed field for models with isothermal background stratification.

Whenever the instability is possible, its nonlinear
development appears to be rather similar for isothermal and adiabatic
background stratification.
In particular, \cite{KBKR11} found that for $\qs=0$, the eigenmode
shows no variation along the direction of the applied magnetic field.
Conversely, if the model is two-dimensional with no extent in the
$y$ direction, which will be assume here, the results are independent
of the value of $\qsz$, so we take in the following $\qsz=0$.
Furthermore, for large magnetic Reynolds number simulations at
different scale separation ratios \citep{KBKMR12} suggest $\qpz=40$
and $\betap=0.05$, which were therefore also the fiducial parameters
used in the study of \cite{KBKR11} and will therefore also be used here.
The results of the DNS presented here suggest somewhat larger values
of $\qpz$ of 130 for Set~A and 95 for Set~B, but this could be a
consequence of intermediate magnetic Reynolds numbers for which
$\qgz$ is known to reach a peak \citep[see Fig.~10 of][]{KBKMR12}.
The dependence on the parameter $\qgz$ has not yet previously been
determined, so this will be done at the end of \Sec{Adia}.
In all other cases we keep $\qgz=0$.
The imposed field strength $B_0$ is chosen such that the minimum of
$\Peff$ occurs near the top boundary and thus $\dd\Peff/\dd\beta^2<0$
in the domain.
We express $B_0$ in units of $\Beqz=\Beq(0)$, which is the equipartition
field strength at $z=0$.

The novel aspect of the present work is that an evolution equation for
the mean specific entropy is included.
Thus, we solve the following system of equations for
the mean vector potential $\meanAA$,
the mean velocity $\meanUU$, the mean density $\meanrho$,
and the mean specific entropy $\means$, in the form
\EQ
{\partial\meanAA\over\partial t}=\meanUU\times\meanv{B}
-\etaT\mu_0\meanJJ,
\EN
\EQ
{\partial\meanUU\over\partial t}=-\meanUU\cdot\nab\meanUU
-\frac{1}{\meanrho}{\bm \nabla}\meanp
+\grav+\meanFFFF^{\rm M}
+\meanFFFF^{\rm K}_{\rm tot},
\label{dUmeandt}
\EN
\EQ
{\partial\meanrho\over\partial t}=-\meanUU\cdot\nab\meanrho
-\meanrho\nab\cdot\meanUU,
\EN
\EQ
\frac{\partial\means}{\partial t}=-\meanUU\cdot\nab\means
-\frac{1}{\meanrho\meanT} \bm{\nabla} \cdot \FF
+2\nuT\meanSSSS^2+\frac{\etat\mu_0}{\meanrho} \overline{\bm{J}}^2
-\frac{1}{\meanT} \Gamma_{\rm cool},\;
\EN
where $\meanv{B}=\BB_0+\nab\times\meanAA$ is the mean
magnetic field including the imposed field,
$\etaT=\etat+\eta$ and $\nuT=\nut+\nu$ are total (turbulent and
microphysical) magnetic diffusivity and viscosity, respectively,
the effective mean Lorentz force is given by Eq.~(\ref{Lor-force}),
and the total viscous force,
$\meanFFFF^{\rm K}_{\rm tot}=(2/\meanrho)\nab\cdot(\meanrho\nuT\meanSSSS)$.
We assume $\nut/\etat=1$ for the {\it turbulent} magnetic Prandtl number.
The mean temperature obeys
$(\gamma-1)\cp\meanT=\gamma\meanp/\meanrho=\cs^2$.
The boundary conditions are stress-free for the velocities, and
perfect conductor boundary conditions for the magnetic field as in
previous mean field models.
In the following we consider two types of
background stratification: isothermal and adiabatic.

\subsection{Isothermal background stratification}
\label{Iso}

We begin by assessing the effects of entropy evolution in the
isothermal models studied by BKKR and \cite{KBKR11}.
In general, the gas will not stay isothermal, because the temperature
changes due to adiabatic expansion and compression.
Indeed, in an isothermally stratified layer a rising blob cools
adiabatically, becomes denser or heavier, and thus experiences a restoring
force with the Brunt-V\"ais\"al\"a frequency $N$, where
$N^2=-\grav\cdot\nab\means/\cp=(\gamma-1)g/\gamma H_\rho$,
with $H_\rho$ being the density scale height.
It turns out that in such a case the negative effective magnetic
pressure instability can be stabilized.
To study this in more detail, we allow for cooling term
of the form $\Gamma_{\rm cool}=(\meanT-T_0)/\cp\meanT\tau$, where
$\meanT$ is the mean temperature, $T_0$ is the reference temperature
of the layer, and $\tau$ is a cooling time.
For the energy flux $\FF$ we assume $\FF=-K \bm{\nabla}\meanT$.

\begin{figure}
\begin{center}
\includegraphics[width=\columnwidth]{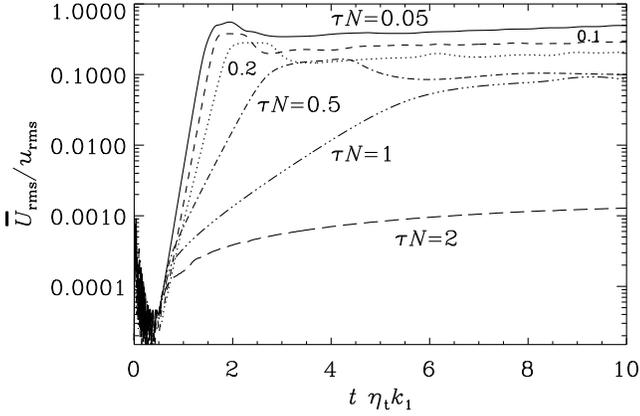}
\end{center}
\caption[]{
Evolution of $\meanU_{\rm rms}/\urms$ for different cooling
times $\tau$ normalized by the Brunt-V\"ais\"al\"a frequency $N$.
}\label{pcomp}
\end{figure}

In \Fig{pcomp} we show the evolution of the rms value of the mean flow
($\meanU_{\rm rms}$) normalized by $\urms=3\kf\etat$ using $\kf=2\pi/H_\rho$,
for different cooling times $\tau$ and $B_0=0.1\Beqz$.
The instability is found to operate only when $\tau N$ is less than
a critical value of order unity.
The largest growth rate seen in \Fig{pcomp} is $\approx60\etat k_1^2$.
The earlier results with an isothermal equation of state are
recovered in the limit $\tau N\to0$, in which case a growth rate
of $\approx110\etat k_1^2$ is found; see Fig.~4 of \cite{KBKR11},
where the growth rate is normalized by $(\nut+\etat)k_1^2$.

\subsection{Adiabatic background stratification}
\label{Adia}

Owing to the stabilizing properties of stable stratification,
we study now the evolution of the instability in an adiabatically
stratified layer, where this stabilizing effect is absent and
the squared sound speed is given by $\cs^2=g(z-\zinfty)$.
Our setup is similar to that of BKR, who considered a reference height
$z=0$ at which initially $\cs=\csz$ and $\rho=\rho_0$,
where $\csz$ and $\rho_0$ are normalization constants.
Length is normalized with respect to the density scale height
$H_{\rho 0}=\cs^2/g$ at $z=0$.
This implies that $\zinfty=(3/2)H_{\rho 0}$.
In BKR the domain extended in the $x$ direction from $-5H_{\rho 0}$
to $+5H_{\rho 0}$, but the resulting horizontal wavelength of the
fastest growing eigenfunction was then about half the $x$ extent.
Therefore we consider here a smaller domain, with
$-3H_{\rho 0}<x<3H_{\rho 0}$ and $-5H_{\rho 0}<z<H_{\rho 0}$.
As in BKR, we choose $B_0/\Beqz=0.01$.
Here we use $\FF=-\chit\meanrho\meanT\bm{\nabla}\means$
for the energy flux
which is appropriate for a turbulent layer, and $\nut/\chit=1$ is the
turbulent Prandtl number,
where $\chit$ is the turbulent heat conductivity.

\begin{figure}
\begin{center}
\includegraphics[width=\columnwidth]{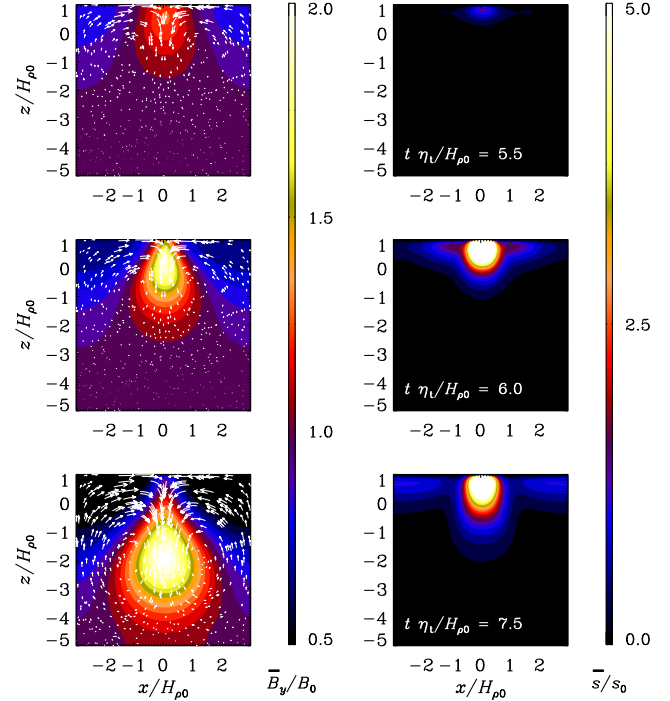}
\end{center}
\caption[]{
Velocity vectors superimposed on color scale representations
of $\meanB_y$ (left) as well as color scale representations of
$\means$ (right) for three different times close to saturation.
Specific entropy is shown in units of $s_0=10^{-4}\cp$.
}\label{pbscomp}
\end{figure}

It turns out that heating is weak, so we ignore the cooling term,
i.e., $\tau\to\infty$.
In \Fig{pbscomp} we show velocity vectors together with $\meanB_y$
as well as $\means$ for three different times close to saturation.
Note that there is a weak enhancement of $\means$ at the location
where the instability develops a positive maximum.
This mean entropy enhancement is associated with turbulent viscous
heating, in particular the contribution $\sim\nuT(\nab\cdot\meanUU)$,
which is important near the surface, even though the magnetic
flux concentration later descends to greater depths.

\begin{figure}
\begin{center}
\includegraphics[width=\columnwidth]{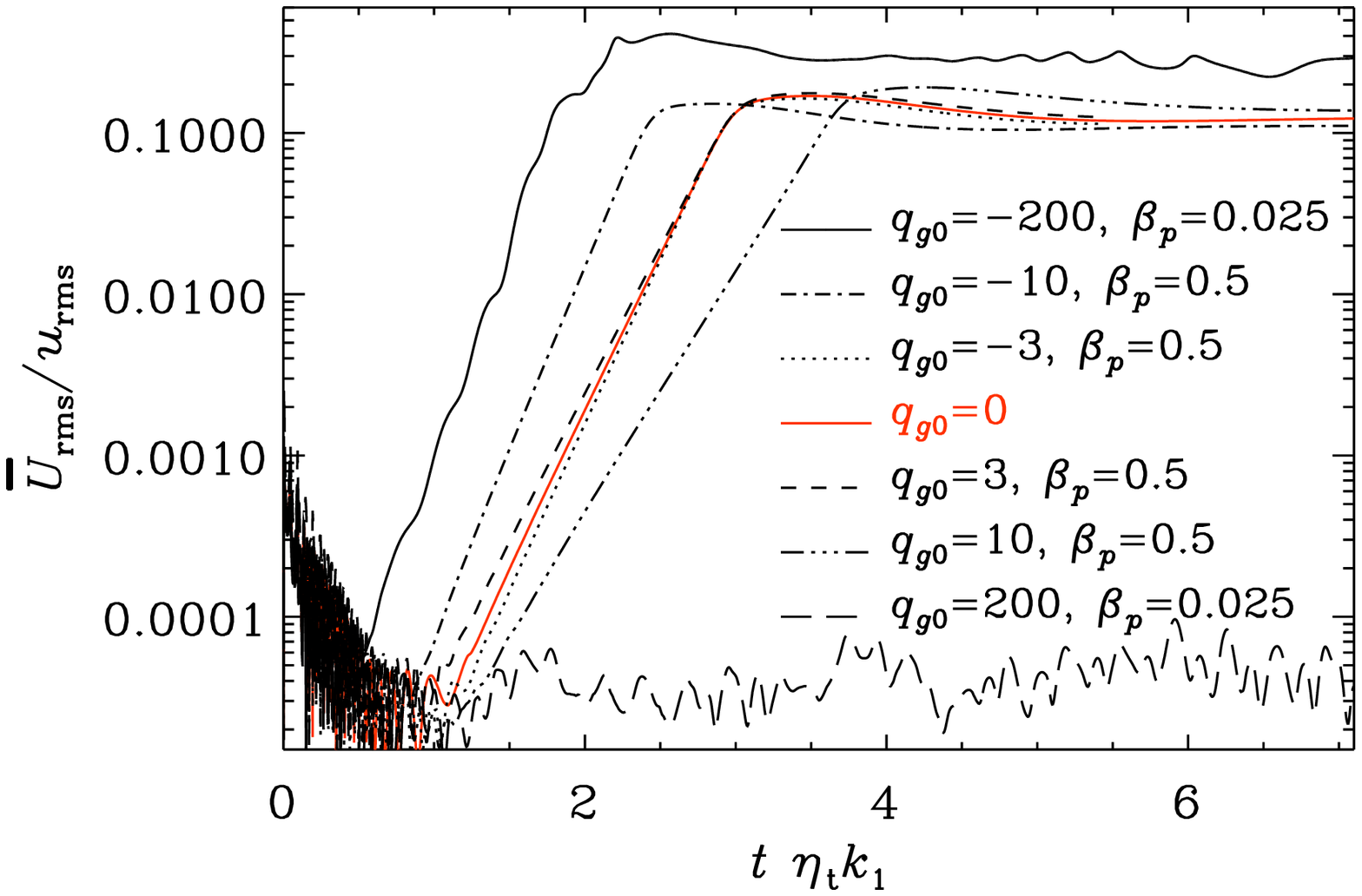}
\end{center}
\caption[]{
Evolution of $\meanU_{\rm rms}/\urms$ for adiabatic stratification
using different combinations of $\qgz$ and $\betag=\Bg/\Beq$.
}\label{pcomp_adia_qg}
\end{figure}

Finally, we use this model to assess the dependence on the parameter $\qgz$,
which can directly contribute to the negative effective magnetic pressure
instability.
According to the DNS we have
$\qgz\approx200$ with $\betag\equiv\Bg/\Beq=0.025$ for Set~A (\Fig{pqeAA1})
and $\qgz\approx3$ with $\betag=0.5$ for Set~B (\Fig{pqeAAA}).
\FFig{pcomp_adia_qg} indicates that the $\qg$ effect would be detrimental to the
instability for Set~A, and negligible for Set~B.
Furthermore, near $\qgz=0$ the dependence of the growth rate on $\qgz$ is
not monotonous: for $\qgz=3$ the growth rate is slightly enhanced and for
$\qgz=10$ it is decreased, but enhanced for $\qgz=-10$ by a similar amount.
The saturation level is only weakly affected by the value of $\qgz$.
We also checked that, as expected from earlier work \citep{KBKR11},
the value of $\qsz$ affects neither the growth rate nor the saturation
value of $\meanU_{\rm rms}$.

\section{Conclusions}

The present simulations have demonstrated that
for weak stratification, and magnetic fields less than the
equipartition value, a destabilising contribution to the mean Lorentz force
is obtained in the presence of turbulent convection.
A similar effect is found for strong density stratification,
although the effect is weaker and limited to a narrower range
in magnetic fields.
Our DNS results agree at least qualitatively with those from
non-stratified (BKR) and stratified (BKKR) forced turbulence
and with theoretical predictions \citep{RK07}, although the minimum
effective magnetic
pressure can now be even more negative and the range where it is
negative extends now to nearly $0.5\Beq$.

Such negative contributions to the effective magnetic pressure
facilitates an instability that can lead to the
generation of flux concentrations from an initially uniform
magnetic field.
This could explain the origin of active regions and sunspots.
However, no clear signs of
instability are found from DNS in turbulent convection.
A possible reason is that the scale
separation in the DNS is insufficient, as has been demonstrated by
\cite{BKKMR11}, who found conclusive evidence of the operation of
the negative effective magnetic pressure instability in DNS
in forced turbulence with a scale separation ratio of 15, using also
spatial-temporal averaging.
For a scale separation ratio of 30, the effect is stronger and flux
concentrations can already be seen without averaging \citep{KBKMR12}.
On the other hand, if the scale separation ratio is as low as 5,
no flux concentrations have been found BKKR.

So far, all studies of the negative effective magnetic pressure effect have
only considered the case of a horizontal mean field.
While this is the most relevant case in view of applications
to stars with differential rotation, the case of a vertical field is
also of interest and might lead to additional effects.
Another important extension of our work is to the case where small-scale
dynamo action becomes important.
This effect is expected to lower the relative importance of the negative
effective
magnetic pressure effect, although this will depend on the value of the
magnetic Prandtl number, which is yet another important parameter whose
effect on the instability is not yet sufficiently well understood.
Finally, we have mentioned the possibility of finite scale separation
effects, which would mean that the scale of the magnetic structure will
be important in determining the efficiency of the negative
effective magnetic pressure effect.
The hope is that such considerations
will provide some insight as to why we have not
yet seen clear evidence for the negative effective
magnetic pressure instability in the present DNS.
Once these various issues are better understood, it becomes timely to
improve mean-field modelling.
Obviously, all the mean-field models of the negative effective
magnetic pressure instability have ignored realistic profiles of
density and turbulent intensity.
Also, the mean-field models should really be three-dimensional
to include the possibility that magnetic structures break up along the
direction of the mean field and form bipolar regions, as was seen
in models of BKR.

\section*{Acknowledgements}

We acknowledge the NORDITA dynamo programs of 2009 and 2011 for
providing a stimulating scientific atmosphere.
The numerical simulations were performed with the supercomputers
hosted by CSC -- IT Center for Science in Espoo, Finland, who are
administered by the Finnish Ministry of Education. Financial support
from the Academy of Finland grant Nos.\ 121431, 136189 (PJK), and
112020 (MJK), the Swedish Research Council grant 621-2007-4064, and
the European Research Council under the AstroDyn Research Project
227952 are acknowledged.


\end{document}